\documentclass[11pt,a4paper, oneside]{extarticle}

\usepackage{amsmath}
\usepackage{amssymb}
\usepackage{mathrsfs}
\usepackage{braket}
\usepackage{bm}

\makeatletter

\@addtoreset{equation}{section}
\makeatother

\usepackage{latexsym}
\def\qed{\\ \hfill $\Box$}
\usepackage{ntheorem}
\theoremstyle{break}
\newtheorem{thm}[subsubsection]{Theorem}

\newtheorem{prop}[subsubsection]{Proposition}
\newtheorem*{thm**}{Theorem}
\newtheorem*{lem**}{Lemma}
\newtheorem*{cor**}{Corollary}
\newtheorem*{prop**}{Proposition}

\theorembodyfont{\normalfont}

\newtheorem*{defi**}{Definition}
\newtheorem*{pf}{Proof}

\theoremheaderfont{\itshape}

\newtheorem*{eg**}{Example}

\usepackage{authblk}
\usepackage{indentfirst}
\usepackage{appendix}
\usepackage{setspace}

\usepackage{graphicx}

\newcommand{\R}{\mathbb{R}}

\newcommand{\ketbra}[2]{\ket{#1}\hspace{-0.25em}\bra{#2}}

\newcommand{\ang}[1]{\left\langle{#1}\right\rangle}

\begin{document}  
\title{Entropic Uncertainty Relations in a Class of Generalized Probabilistic Theories}

\author{Ryo Takakura\thanks{takakura.ryo.27v@st.kyoto-u.ac.jp},\quad Takayuki Miyadera\thanks{miyadera@nucleng.kyoto-u.ac.jp}}
\affil{Department of Nuclear Engineering 
	\\Kyoto University
	\\Kyoto daigaku-katsura, Nishikyo-ku, Kyoto, 615-8540, Japan}
\date{}
\maketitle

\begin{abstract}
Entropic uncertainty relations play an important role in both fundamentals and applications of quantum theory. 
Although they have been well-investigated in quantum theory, little is known about entropic uncertainty in generalized probabilistic theories (GPTs). 
The current study explores two types of entropic uncertainty relations, preparation and measurement uncertainty relations, in a class of GPTs which can be considered generalizations of quantum theory.  
Not only a method for obtaining entropic preparation uncertainty relations but also an entropic measurement uncertainty relation similar to the quantum one by Buscemi {\it et al.} [Phys. Rev. Lett., {\bf 112}, 050401] are proved in those theories.
It manifests that the entropic structure of uncertainty relations in quantum theory is more universal.
Concrete calculations of our relations in GPTs called the regular polygon theories are also demonstrated.
\end{abstract}	



\section{Introduction}
The concept of uncertainty, advocated initially by Heisenberg \cite{Heisenberg1927}, is one of the most peculiar features in quantum theory. 
Much study has been devoted to proper understandings of uncertainty to demonstrate that it has two aspects: {\it preparation uncertainty} and {\it measurement uncertainty} \cite{Busch_quantummeasurement}.
Loosely speaking, for a pair of observables which do not commute, the former describes that there is no state on which their individual measurements output simultaneously definite values, while the latter expresses the impossibility of performing their joint measurement. 
Although there have been considered several mathematical representations of them, {\it preparation uncertainty relations} ({\it PURs}) and {\it measurement uncertainty relations} ({\it MURs}) respectively \cite{PhysRev.34.163,PhysRevLett.60.2447,BUSCH2007155,doi:10.1063/1.4871444,PhysRevA.67.042105}, entropic uncertainty relations \cite{UncertaintyRelationsForInformationEntropy,PhysRevLett.50.631,PhysRevLett.60.1103,10.2307/25051432,PhysRevLett.112.050401} have the advantages of their compatibility with information theory and independence from the structure of the sample spaces. They indeed have been applied to the field of quantum information in various ways \cite{RevModPhys.89.015002}. 
On the other hand, two kinds of uncertainty have been investigated also in physical theories broader than quantum theory called {\it generalized probabilistic theories} ({\it GPTs}) \cite{Gudder_stochastic,hardy2001quantum,PhysRevA.75.032304,PhysRevA.84.012311,BARNUM20113,1751-8121-47-32-323001}. For example, there have been researches on both types of uncertainty \cite{PhysRevA.101.052104} or joint measurability of observables \cite{PhysRevLett.103.230402,Busch_2013,PhysRevA.89.022123,PhysRevA.94.042108,PhysRevA.98.012133}, which are related with MURs, in GPTs.
In \cite{takakura2020uncertainty}, several formulations of two types of uncertainty were generalized to GPTs, and it was revealed quantitatively that there are close relations between them not only in quantum theory \cite{doi:10.1063/1.3614503} but also in a class of GPTs. 
However, although the notion of entropy has been introduced in GPTs \cite{1367-2630-12-3-033024,Short_2010,KIMURA2010175,EPTCS195.4,1367-2630-19-4-043025,Takakura_2019}, insights of entropic uncertainty relations in GPTs are still missing.

In the present paper, entropic uncertainty relations are studied in a class of GPTs investigated in the previous work \cite{takakura2020uncertainty}: GPTs satisfying {\it transitivity} and {\it self-duality} with respect to a certain inner product. 
They include finite dimensional classical and quantum theories, and thus can be regarded as generalizations of them. In those theories, we obtain an entropic inequality related with PURs in a simple way via the Landau-Pollak-type relations \cite{Uffink_PhD,PhysRevA.71.052325,PhysRevA.76.062108}. We also prove an entropic relation similar to the quantum MUR by Buscemi {\it et al.} \cite{PhysRevLett.112.050401} with their formulations generalized to those GPTs.
Our results manifest that the structures of entropic PURs and MURs in quantum theory are indeed more universal ones.
Moreover, they can be considered as an entropic counterpart of \cite{takakura2020uncertainty}: if there exist entropic PURs giving certain bounds of uncertainty, then entropic MURs also exist and can be formulated in terms of the same bounds as PURs.
We also present, as an illustration, concrete expressions of our entropic relations in a specific class of GPTs called the {\it regular polygon theories} \cite{1367-2630-13-6-063024}.

This paper is organized as follows. In section \ref{sec:GPTs}, we give a short survey of GPTs including the introduction of the regular polygon theories. Section \ref{sec:main section} is the main part of this paper, and there are shown entropic uncertainty relations in a certain class of GPTs. We conclude the present work and give brief discussions in section \ref{sec:conclusion}.

\section{GPTs}
\label{sec:GPTs}
GPTs are the most general physical theories reflecting intuitively the notion of physical experiments: to prepare a state, to conduct a measurement, and to observe a probability distribution. In this section, a brief survey of GPTs is shown according mainly to \cite{1751-8121-47-32-323001,takakura2020uncertainty,KIMURA2010175,EPTCS195.4,kimura2010physical}.

\subsection{Fundamentals}
\label{subsec:fundamentals}
Any GPT is associated with the notion of {\it states} and {\it effects}. In this paper, a compact convex set $\Omega$ in $V\equiv\R^{N+1}$ with $\mathrm{dim}\mathit{aff}(\Omega)=N$ describes the set of all states in a GPT, which we call the {\it state space} of the theory. We assume in this paper that GPTs are finite dimensional ($N<\infty$), and $\mathit{aff}(\Omega)$ does not include the origin $O$ of $V$ ($O\notin\mathit{aff}(\Omega)$). We note that the notion of probability mixture of states is reflected by the convex structure of $\Omega$. 
The extreme elements of $\Omega$ are called {\it pure states}, and we denote the set of all pure states by $\Omega^{\mathrm{ext}}=\{\omega^{\mathrm{ext}}_{\lambda}\}_{\lambda\in\Lambda}$. 
The other elements of $\Omega$ are called {\it mixed states}.
For a GPT with its state space $\Omega$, we define the {\it effect space} of the theory as $\mathcal{E}(\Omega)=\{e\in V^{*}\mid e(\omega)\in[0, 1]\ \mbox{for all}\ \omega\in\Omega\}$, where $V^{*}$ is the dual space of $V$, 
and call its elements {\it effects}.
Remark that we follow the {\it no-restriction hypothesis} \cite{PhysRevA.81.062348} in this paper, and we sometimes denote $\mathcal{E}(\Omega)$ simply by $\mathcal{E}$. Introducing the {\it unit effect} $u$ as $u\in\mathcal{E}(\Omega)$ satisfying $u(\omega)=1$ for all $\omega\in\Omega$, a {\it measurement} or {\it observable} on some sample space $X$ is defined by a set of effects $\{e_{x}\}_{x\in X}$ such that $\sum_{x\in X}e_{x}=u$. In this paper, we assume that every measurement is with finite outcomes (i.e. the sample space $X$ is finite) and does not include the zero effect, and the trivial measurement $\{u\}$ is not considered. Two measurements $A=\{a_{x}\}_{x\in X}$ and $B=\{b_{y}\}_{y\in Y}$ are called {\it jointly measurable} or {\it compatible} if there exists a joint measurement $C=\{c_{xy}\}_{(x, y)\in X\times Y}$ such that its marginals satisfy $\sum_{y\in Y}c_{xy}=a_{x}$ and $\sum_{x\in X}c_{xy}=b_{y}$ for all $x\in X, y\in Y$. If $A$ and $B$ are not jointly measurable, then they are called {\it incompatible}. We say that two GPTs are equivalent if their state spaces $\Omega_{1}$ and $\Omega_{2}$ satisfies $\psi(\Omega_{1})=\Omega_{2}$ for a linear bijection $\psi$ on $V$.
In that case, because $\mathcal{E}(\Omega_{2})=\mathcal{E}(\Omega_{1})\circ\psi^{-1}$ holds, we can see the covariance (equivalence) of physical predictions.

For a state space $\Omega$, the {\it positive cone} $V_{+}(\Omega)$ (or simply $V_{+}$) generated by $\Omega$ is defined as the set of all unnormalized states, that is, $V_{+}:=\{v\in V\mid v=k\omega, \omega\in\Omega, k\ge0\}$. We can also define the {\it dual cone} $V^{*}_{+}(\Omega)$ (or simply $V^{*}_{+}$) as the set of all unnormalized effects: $V^{*}_{+}:=\{f\in V^{*}\mid f(v)\ge0, \ ^\forall v\in V_{+}\}$. 
A half-line $E\subset V_{+}$ is called an {\it extremal ray} of $V_{+}$ (respectively $V_{+}^{*}$) if $l=m+n$ with $l\in E$ and $m, n\in V_{+}$ (respectively $m, n\in V_{+}^{*}$) implies $m, n\in E$. We call effects on extremal rays of $V_{+}^{*}$ {\it indecomposable}, while it is easy to see that the half-lines $\{x\in V\mid x=k\omega^{\mathrm{ext}}_{\lambda}, k>0\}$ generated by the pure states $\Omega^{\mathrm{ext}}=\{\omega^{\mathrm{ext}}_{\lambda}\}_{\lambda\in \Lambda}$ are the extremal rays of $V_{+}$. It is known that there exist pure and indecomposable effects, and we denote by $\mathcal{E}^{\mathrm{ext}}(\Omega)$ (or simply $\mathcal{E}^{\mathrm{ext}}$) the set of all pure and indecomposable effects. They are thought to be a generalization of rank-1 projections in finite dimensional quantum theories (see \cite{takakura2020uncertainty}).

\subsection{Additional notions}
\label{subsec:additional notions}
Let $\Omega\subset V$ be a state space.
A linear bijection $T\colon V\to V$ is called a {\it state automorphism} on $\Omega$ if it satisfies $T(\Omega)=\Omega$, and we denote by $G(\Omega)$ (or simply $G$) the set of all state automorphisms on $\Omega$. 
States $\omega_{1}, \omega_{2}\in\Omega$ are called {\it physically equivalent} if there exists $T\in G$ satisfying $T\omega_{1}=\omega_{2}$.
We say that $\Omega$ is {\it transitive} if all pure states are physically equivalent, i.e. for an arbitrary pair of pure states $\omega^{\mathrm{ext}}_{i}, \omega^{\mathrm{ext}}_{j}\in\Omega^{\mathrm{ext}}$ there exists $T\in G$ such that $T\omega^{\mathrm{ext}}_{i}=\omega^{\mathrm{ext}}_{j}$. 
When $\Omega$ is transitive, we can define the {\it maximally mixed state} $\omega_{M}\in\Omega$ as a unique state satisfying $T\omega_{M}=\omega_{M}$ for all $T\in G$ \cite{Davies_compactconvex}.

There exists a useful inner product $\langle\cdot ,\cdot\rangle_{G}$ on $V$, with respect to which all elements of $G$ are orthogonal transformations on $V$.
That is,
\[
\ang{Tx, Ty}_{G}=\ang{x, y}_{G}\quad\ ^{\forall}x, y\in V
\]
holds for all $T\in G$. In fact, $\ang{\cdot, \cdot}_{G}$ is constructed in the way
\[
\ang{x, y}_{G}=\int_{G}(x, y)d\mu\quad (x, y\in V)
\]
by means of the two-sided invariant Haar measure $\mu$ on $G$ and a reference inner product $(\cdot,\cdot)$ on $V$ (such as the standard Euclidean inner product on $V$).
When $\Omega$ is transitive, we can prove that all pure states are of equal norm with respect to $\langle\cdot,\cdot\rangle_{G}$\ :
\begin{align}
\label{eq:equal norm}
\|\omega_{\lambda}^{\mathrm{ext}}\|_{G}=\sqrt{\alpha} \quad ^\forall \omega_{\lambda}^{\mathrm{ext}}\in\Omega^{\mathrm{ext}},
\end{align}
where $\|\cdot\|_{G}:=\langle\cdot,\cdot\rangle_{G}^{1/2}$ and $\alpha$ is a positive number.

For the positive cone $V_{+}$ generated by $\Omega$ and an inner product $(\cdot,\cdot)$ on $V$, the {\it internal dual cone} $V_{+(\cdot,\cdot)}^{*int}$ relative to $(\cdot,\cdot)$ is defined as $V_{+(\cdot,\cdot )}^{*int}:=\{w\in V\mid (w, v)\ge0\ ^{\forall}v\in V_{+}\}$, and the cone $V_{+}$ is called {\it self-dual} if $V_{+}=V_{+(\cdot ,\cdot)}^{*int}$ for some inner product $(\cdot,\cdot)$ on $V$. Note that by virtue of the Riesz representation theorem $V_{+(\cdot,\cdot)}^{*int}$ can be regarded as the dual cone $V^{*}_{+}$, i.e. the set of all unnormalized effects. Thus, the self-duality of $V_{+}$ means that (unnormalized) states can be identified with (unnormalized) effects. Let us assume that $\Omega$ is transitive and $V_{+}$ is self-dual with the self-dualizing inner product being $\langle\cdot,\cdot\rangle_{G}$. In this case, \eqref{eq:equal norm} holds, and we can prove that $\alpha\mathcal{E}^{\mathrm{ext}}=\Omega^{\mathrm{ext}}$, that is, 
\begin{align}
\label{def:pure and indecomp effect}
e_{\lambda}^{\mathrm{ext}}:=\frac{\omega_{\lambda}^{\mathrm{ext}}}{\alpha}
\end{align}
gives a pure and indecomposable effect for any $\omega_{\lambda}^{\mathrm{ext}}\in\Omega^{\mathrm{ext}}$ because the extreme rays of $V_{+}=V_{+\langle\ ,\ \rangle_{G}}^{*int}$ are generated by $\Omega^{\mathrm{ext}}$ and $\omega_{\lambda}^{\mathrm{ext}}$ is a (unique) state satisfying $\langle e_{\lambda}^{\mathrm{ext}}, \omega_{\lambda}^{\mathrm{ext}}\rangle_{G}=1$ \cite{KIMURA2010175}. In \cite{takakura2020uncertainty}, it was demonstrated that the state spaces of finite dimensional classical and quantum theories satisfy both transitivity and self-duality with respect to the inner product $\langle\cdot,\cdot\rangle_{G}$. There was also shown a proposition about the self-duality with respect to $\langle\cdot,\cdot\rangle_{G}$ in \cite{takakura2020uncertainty}. 
\begin{prop}[Proposition 2.3.2 in \cite{takakura2020uncertainty}]
	\label{prop_self-duality}
	Let $\Omega$ be transitive with $|\Omega^{\mathrm{ext}}|<\infty$ and $V_+$ be self-dual with respect to some inner product. There exists a linear bijection $\Xi\colon V\to V$ such that $\Omega':=\Xi\Omega$ is transitive and the generating positive cone $V'_{+}$ is self-dual with respect to $\langle\cdot,\cdot\rangle_{G(\Omega')}$, i.e.
	$V^{'}_+ = V_{+\langle\cdot ,\cdot \rangle_{G(\Omega')}}^{'*int}$.
\end{prop}
Proposition \ref{prop_self-duality} demonstrates that 
when $\Omega$ is transitive and $\Omega^{\mathrm{ext}}$ is finite, we can find another expression $\Omega'$ of the theory which is transitive and whose positive cone is self-dual with respect to $\langle\cdot ,\cdot \rangle_{G(\Omega')}$. 

\subsection{Examples of GPTs: regular polygon theories}
\label{subsec:polygons}
The {\it regular polygon theories} are GPTs whose state spaces are the regular polygons in $V\equiv\R^{3}$, and if a state space is the regular polygon with $n$ sides ($n\ge3$), then we denote it by $\Omega_{n}$. In \cite{1367-2630-13-6-063024}, we can find that $\Omega_{n}$ is given by the convex hull of its pure states (its vertices)
\begin{align}
\label{def:polygon pure state0}
\Omega^{\mathrm{ext}}_{n}=\{\omega_{n}^{\mathrm{ext}}(i)\}_{i=0}
^{n-1},
\end{align}
where
\begin{align}
\label{def:polygon pure state}
\omega_{n}^{\mathrm{ext}}(i)=
\left(
\begin{array}{c}
r_{n}\cos({\frac{2\pi i}{n}})\\
r_{n}\sin({\frac{2\pi i}{n}})\\
1
\end{array}
\right)\quad\mbox{with}\quad r_{n}=\sqrt{\frac{1}{\cos({\frac{\pi}{n}})}}.
\end{align}
The corresponding effect space $\mathcal{E}(\Omega_{n})$ is given by $V^{*int}_{+(\ ,\ )_{E}}\cap\{u-V^{*int}_{+(\ ,\ )_{E}}\}$ in terms of the dual cone $V^{*}_{+}(\Omega_{n})=V^{*int}_{+(\ ,\ )_{E}}$ represented by the standard Euclidean inner product $(\ ,\ )_{E}$ of $V$. $V^{*}_{+}(\Omega_{n})=V^{*int}_{+(\ ,\ )_{E}}$ is generated by the pure and indecomposable effects
\begin{align}
\label{def:polygon pure effect0}
\mathcal{E}^{\mathrm{ext}}(\Omega_{n})=\{e_{n}^{\mathrm{ext}}(i)\}_{i=0}
^{n-1},
\end{align}
where
\begin{equation}
\label{def:polygon pure effect}
\begin{aligned}
&e_{n}^{\mathrm{ext}}(i)=\frac{1}{2}
\left(
\begin{array}{c}
r_{n}\cos({\frac{(2i-1)\pi}{n}})\\
r_{n}\sin({\frac{(2i-1)\pi}{n}})\\
1
\end{array}
\right)\ \ (n:\mbox{even})\ ;\\
&e_{n}^{\mathrm{ext}}(i)=\frac{1}{1+r_{n}^{2}}
\left(
\begin{array}{c}
r_{n}\cos({\frac{2i\pi}{n}})\\
r_{n}\sin({\frac{2i\pi}{n}})\\
1
\end{array}
\right)\ \ (n:\mbox{odd}).
\end{aligned}
\end{equation}
We can also consider the case when $n=\infty$ in \eqref{def:polygon pure state} - \eqref{def:polygon pure effect}. The state space $\Omega_{\infty}$ is a disc with its pure states and pure and indecomposable effects being
\begin{equation}
\label{def:disc pure state0}
\Omega_{\infty}^{\mathrm{ext}}=\{\omega_{\infty}^{\mathrm{ext}}(\theta)\}_{\theta\in[0, 2\pi)}
\end{equation}
and
\begin{equation}
\label{def:disc pure effect0}
\mathcal{E}^{\mathrm{ext}}(\Omega_{\infty})=\{e_{\infty}^{\mathrm{ext}}(\theta)\}_{\theta\in[0, 2\pi)},
\end{equation}
where
\begin{align}
\label{def:disc pure state and effect}
\omega_{\infty}^{\mathrm{ext}}(\theta)=
\left(
\begin{array}{c}
\cos\theta\\
\sin\theta\\
1
\end{array}
\right)\quad\mbox{and}\quad
e_{\infty}^{\mathrm{ext}}(\theta)=\frac{1}{2}
\left(
\begin{array}{c}
\cos\theta\\
\sin\theta\\
1
\end{array}
\right)
\end{align}
respectively. 

For $n=3, 4, \cdots, \infty$, it can be shown that $\Omega_{n}$ is transitive with respect to $G(\Omega_{n})$, and the standard Euclidean inner product $(\cdot,\cdot)_{E}$ is indeed the inner product $\langle\cdot,\cdot\rangle_{G(\Omega_{n})}$ invariant with any $T\in G(\Omega_{n})$. Moreover, we can see from \eqref{def:polygon pure state0} - \eqref{def:disc pure state and effect} that $V_{+}(\Omega_{n})$ is self-dual with respect to $(\cdot ,\cdot )_{E}=\langle\cdot ,\cdot \rangle_{G(\Omega_{n})}$ when $n$ is odd or $\infty$, whereas $V_{+}(\Omega_{n})$ is no more than isomorphic to $V_{+ \langle\cdot ,\cdot \rangle_{G(\Omega_{n})}}^{*int}$ when $n$ is even (in this case, $V_{+}(\Omega_{n})$ is called {\it weakly self-dual} \cite{barnum2012teleportation,Barnum2013}). We note that the cases when $n=3$ and $n=\infty$ correspond to a classical trit system and a qubit system restricted to real coefficients respectively.

\section{Entropic Uncertainty Relations in a class of GPTs}
\label{sec:main section}
In this section, we present our main results on two types of entropic uncertainty in a certain class of GPTs. While our results reproduce entropic uncertainty relations obtained in finite dimensional quantum theories, they indicate that similar relations hold also in a broader class of physical theories. We also demonstrate entropic uncertainty relations in the regular polygon theories as an illustration of our results.

\subsection{Entropic PURs}
\label{subsec:entropic PURs}
In quantum theory, it is known that we cannot prepare a state on which individual measurements of position and momentum observables, for example, take simultaneously definite values \cite{Kennard1927}. This type of uncertainty and its quantifications are called {\it preparation uncertainty} and {\it preparation uncertainty relations} ({\it PURs}) respectively.

In order to give general descriptions of uncertainty in GPTs, the notion of ideal measurements has to be introduced. Considering that projection-valued measures (PVMs), whose effects are sums of rank-1 projections, give ideal measurements in finite dimensional quantum theories \cite{Busch_quantummeasurement}, we call a measurement $\{e_{x}\}_{x\in X}$ in some GPT {\it ideal} \cite{takakura2020uncertainty} if for any $x\in X$ there exists a finite set of pure and indecomposable effects $\{e^{\mathrm{ext}}_{i_{x}}\}_{i_{x}}$ such that
\begin{align}
\label{def:ideal measurement}
e_{x}=\sum_{i_{x}}e^{\mathrm{ext}}_{i_{x}} \quad\mbox{or}\quad e_{x}=u-\sum_{i_{x}}e^{\mathrm{ext}}_{i_{x}}.
\end{align}
It is easy to check that measurements satisfying $\eqref{def:ideal measurement}$ are reduced to PVMs in finite dimensional quantum theories.

Let us consider a GPT with its state space $\Omega$, and two ideal measurements $A=\{a_{x}\}_{x\in X}$ and $B=\{b_{y}\}_{y\in Y}$ on $\Omega$. For the probability distribution $\{a_{x}(\omega)\}_{x}$ obtained in the measurement of $A$ on a state $\omega\in\Omega$  (and similarly for $\{b_{y}(\omega)\}_{y}$), its Shannon entropy is defined as 
\begin{align}
	\label{def:Shannon entropy}
	H\left(\{a_{x}(\omega)\}_{x}\right)=-\sum_{x\in X}a_{x}(\omega)\log{a_{x}(\omega)}.
\end{align}
One way to obtain an entropic PUR is to consider the Landau-Pollak-type relations \cite{Uffink_PhD,PhysRevA.71.052325,PhysRevA.76.062108}:
\begin{align}
\label{def:L-P UR}
\max_{x\in X}a_{x}(\omega)+\max_{y\in Y}b_{y}(\omega)\le \gamma_{A,B}\quad\ ^\forall\omega\in\Omega
\end{align}
with a constant $\gamma_{A,B}\in(0, 2]$. Remark that relations of the form \eqref{def:L-P UR} always can be found for any pair of measurements. It is known \cite{PhysRevLett.60.1103, inequalities1988} that $\max_{x\in X}a_{x}(\omega)$ is related with $H\left(\{a_{x}(\omega)\}_{x}\right)$ by
\[
\exp\left[-H\left(\{a_{x}(\omega)\}_{x}\right)\right]\le\max_{x\in X}a_{x}(\omega),
\] 
and thus we can observe from \eqref{def:L-P UR}
\begin{align*}
\exp\left[-H\left(\{a_{x}(\omega)\}_{x}\right)\right]+\exp\left[-H\left(\{b_{y}(\omega)\}_{y}\right)\right]\le \gamma_{A, B}.
\end{align*}
Considering that
\begin{align*}
&\exp\left[-H\left(\{a_{x}(\omega)\}_{x}\right)\right]+\exp\left[-H\left(\{b_{y}(\omega)\}_{y}\right)\right]\\ 
&\qquad\qquad\qquad\qquad\qquad\qquad\ge 2\exp\left[\frac{-H\left(\{a_{x}(\omega)\}_{x}\right)-H\left(\{b_{y}(\omega)\}_{y}\right)}{2}\right]
\end{align*} 
holds, we can finally obtain an entropic relation
\begin{align}
\label{eq:entropic PUR via L-P}
H\left(\{a_{x}(\omega)\}_{x}\right)+H\left(\{b_{y}(\omega)\}_{y}\right)\ge -2\log\frac{\gamma_{A, B}}{2}\quad\ ^\forall\omega\in\Omega.
\end{align}
If $\gamma_{A, B}<2$, then \eqref{eq:entropic PUR via L-P} gives an entropic PUR because it indicates that it is impossible to prepare a state which makes both $H\left(\{a_{x}(\omega)\}_{x}\right)$ and $H\left(\{b_{y}(\omega)\}_{y}\right)$ zero, that is, there is no state preparation on which $A$ and $B$ take simultaneously definite values (note that \eqref{def:L-P UR} also gives a PUR if $\gamma_{A,B}<2$). In a finite dimensional quantum theory with its state space $\Omega_{\mathrm{QT}}$, it can be shown that 
\begin{align}
\label{eq:quantum LP}
\max_{x}a_{x}(\omega)+\max_{y}b_{y}(\omega)\le1+\max_{x, y}|\braket{a_{x}|b_{y}}|\quad\ ^\forall\omega\in\Omega_{\mathrm{QT}},
\end{align}
where $A=\{\ketbra{a_{x}}{a_{x}}\}_{x}$ and $B=\{\ketbra{b_{y}}{b_{y}}\}_{y}$ are rank-1 PVMs. In that case, \eqref{eq:entropic PUR via L-P} can be rewritten as
\begin{align}
\label{eq:Deutsch ent PUR}
H\left(\{a_{x}(\omega)\}_{x}\right)+H\left(\{b_{y}(\omega)\}_{y}\right)\ge2\log\frac{2}{1+\underset{x, y}{\max}|\braket{a_{x}|b_{y}}|}\quad\ ^\forall\omega\in\Omega_{\mathrm{QT}},
\end{align}
which is the entropic PUR proved by Deutsch \cite{PhysRevLett.50.631}. There have been studies to find a better bound \cite{PhysRevLett.60.1103} or generalization \cite{10.2307/25051432} of \eqref{eq:Deutsch ent PUR}.

\subsection{Entropic MURs}
\label{subsec:entropic MURs}
When considering two measurements, they are not always jointly measurable \cite{PhysRevA.94.042108}. Their incompatibility is represented quantitatively by {\it measurement uncertainty relations} ({\it MURs}) in terms of {\it measurement error}, which describes the difference between the ideal, original measurement and their {\it approximate joint measurement} \cite{Busch_2013,PhysRevA.89.022123}.

Let $\Omega$ be a state space which is transitive and satisfies $V_{+}(\Omega)\equiv V_{+}=V^{*int}_{+\langle\cdot,\cdot\rangle_{G}}$, and we hereafter denote the inner product $\langle\cdot,\cdot\rangle_{G}$ simply by $\langle\cdot,\cdot\rangle$. Then, because of the self-duality of $V_{+}$, we can in the following identify effects with elements of $V_{+}$. There can be defined measurement error in terms of entropy in the identical way with the quantum one by Buscemi {\it et al.} \cite{PhysRevLett.112.050401}. Let in the GPT $E=\{e_{x}\}_{x\in X}$ be an ideal measurement defined in \eqref{def:ideal measurement} and $\mathcal{M}=\{m_{\hat{x}}\}_{\hat{x}\in \hat{X}}$ be a measurement. Since it was demonstrated in \cite{takakura2020uncertainty} that
\begin{align}
\label{eq:eigenstate}
\left\langle e_{x'}, \frac{e_{x}}{\langle u, e_{x}\rangle}\right\rangle=\delta_{x'x}
\end{align}
holds for all $x, x'\in X$, and
\begin{equation}
\begin{aligned}
\omega_{M}=u&=\sum_{x}e_{x}\\
&=\sum_{x}\langle u, e_{x}\rangle \frac{e_{x}}{\langle u, e_{x}\rangle}
\end{aligned}
\end{equation}
holds, the joint probability distribution 
\begin{equation}
\label{def:joint dist}
\{p(x, \hat{x})\}_{x, \hat{x}}=\{\langle e_{x}, m_{\hat{x}}\rangle\}_{x, \hat{x}}=\left\{\langle u, e_{x}\rangle\left\langle\frac{e_{x}}{\langle u, e_{x}\rangle}, m_{\hat{x}}\right\rangle\right\}_{x, \hat{x}}
\end{equation}
is considered to be obtained in the measurement of $\mathcal{M}$ on the eigenstates $\{e_{x}/\langle u, e_{x} \rangle\}_{x}$ of $E$ (see \eqref{eq:eigenstate}) with the initial distribution
\begin{align}
\label{eq:initial dist}
\left\{
p(x)
\right\}_{x}=
\left\{
\langle u, e_{x}\rangle
\right\}_{x}.
\end{align}
In fact, as shown in \cite{PhysRevLett.112.050401}, the conditional entropy
\begin{equation}
\label{def:entropic MN}
\begin{aligned}
\mathsf{N}(\mathcal{M};E):
&=H(E|\mathcal{M})\\
&=\sum_{\hat{x}}p(\hat{x})H\left(\{p(x|\hat{x})\}_{x}\right)\\
&=\sum_{\hat{x}}\ang{u, m_{\hat{x}}}H\left(\left\{\ang{e_{x}, \frac{m_{\hat{x}}}{\ang{u, m_{\hat{x}}}}}\right\}_{x}\right)
\end{aligned}
\end{equation}
calculated via \eqref{def:joint dist} describes how inaccurately the actual measurement $\mathcal{M}$ can estimate the input eigenstates of the ideal measurement $E$. Strictly speaking, if we consider measuring $\mathcal{M}$ on $e_{x}/\langle u, e_{x}\rangle$ and estimating the input state from the output probability distribution 
\[
\{p(\hat{x}|x)\}_{\hat{x}}=\left\{\left\langle m_{\hat{x}}, \frac{e_{x}}{\langle u, e_x\rangle}\right\rangle\right\}_{\hat{x}}
\]
by means of a guessing function $f:\hat{X}\to X$, then the error probability $p_{\mathrm{error}}^{f}(x)$ is given by
\[
p_{\mathrm{error}}^{f}(x)=1-\sum_{\hat{x}: f(\hat{x})=x}p(\hat{x}|x)=\sum_{\hat{x}: f(\hat{x})\neq x}p(\hat{x}|x).
\]
When similar procedures are conducted for all $x\in X$ with the probability distribution $\{p(x)\}_{x}$ in \eqref{eq:initial dist}, the total error probability $p_{\mathrm{error}}^{f}$ is
\begin{align}
p_{\mathrm{error}}^{f}=\sum_{x}p(x)\ p_{\mathrm{error}}^{f}(x)=\sum_{x\in X}\sum_{\hat{x}: f(\hat{x})\neq x}p(x, \hat{x}),
\end{align}
and it was shown in \cite{PhysRevLett.112.050401} that 
\[
\min_{f}p_{\mathrm{error}}^{f}\to0\quad\iff\quad\mathsf{N}(\mathcal{M};E)=H(E|\mathcal{M})\to0.
\]
We can conclude from the consideration above that the entropic quantity \eqref{def:entropic MN} represents the difference between $E$ to be measured ideally and $\mathcal{M}$ measured actually, and thus we can define their entropic measurement error as \eqref{def:entropic MN}.

We are now in the position to derive a similar entropic relation to \cite{PhysRevLett.112.050401} with the generalized entropic measurement error \eqref{def:entropic MN}. We continue focusing on a GPT with its state space $\Omega$ being transitive and $V_{+}$ being self-dual with respect to the inner product $\langle\cdot,\cdot\rangle_{G}\equiv\langle\cdot,\cdot\rangle$, that is, $V_{+}=V^{*int}_{+\langle\cdot,\cdot\rangle}$. Let $A=\{a_{x}\}_{x\in X}$ and $B=\{b_{y}\}_{y\in Y}$ be a pair of ideal measurements defined in \eqref{def:ideal measurement}, and consider their approximate joint measurement $\mathcal{M}=\{m_{\hat{x}\hat{y}}\}_{(\hat{x},\hat{y})\in X\times Y}$ and its marginals
\begin{align*}
&\mathcal{M}^{A}=\{m_{\hat{x}}\}_{\hat{x}\in X}\quad\mbox{with}\quad m_{\hat{x}}=\sum_{\hat{y}\in Y}m_{\hat{x}\hat{y}}\\
&\mathcal{M}^{B}=\{m_{\hat{y}}\}_{\hat{y}\in Y}\quad\mbox{with}\quad m_{\hat{y}}=\sum_{\hat{x}\in X}m_{\hat{x}\hat{y}}.
\end{align*}
We can prove the following theorem.
\begin{thm}
\label{thm:entropic MUR}
Suppose that $\Omega$ is a transitive state space with its positive cone $V_{+}$ being self-dual with respect to $\langle\cdot,\cdot\rangle_{G}\equiv\langle\cdot,\cdot\rangle$, $A=\{a_{x}\}_{x}$ and $B=\{b_{y}\}_{y}$ are ideal measurements on $\Omega$, and $\mathcal{M}$ is an arbitrary approximate joint measurement of $(A, B)$ with its marginals $\mathcal{M}^{A}$ and $\mathcal{M}^{B}$.
If there exists a relation
\begin{align*}
H\left(\{a_{x}(\omega)\}_{x}\right)+H\left(\{b_{y}(\omega)\}_{y}\right)\ge \Gamma_{A, B}\quad\ ^\forall\omega\in\Omega
\end{align*}
with a constant $\Gamma_{A, B}$, then it also holds that
\begin{align*}
\mathsf{N}(\mathcal{M}^{A};A)+\mathsf{N}(\mathcal{M}^{B};B)\ge \Gamma_{A, B}.
\end{align*}
\end{thm}
\begin{pf}
Since for every $\hat{x}\in X$ and $\hat{y}\in Y$ $\omega_{\hat{x}\hat{y}}:=m_{\hat{x}\hat{y}}/\langle u, m_{\hat{x}\hat{y}}\rangle$ is a state, it holds that
\[
H\left(\{a_{x}(\omega_{\hat{x}\hat{y}})\}_{x}\right)+H\left(\{b_{y}(\omega_{\hat{x}\hat{y}})\}_{y}\right)\ge \Gamma_{A, B}
\]
for all $\hat{x}\in X$ and $\hat{y}\in Y$. Therefore, taking into consideration that $\langle u, m_{\hat{x}\hat{y}}\rangle\ge0$ for all $\hat{x}, \hat{y}$ and $\sum_{\hat{x}\hat{y}}\langle u, m_{\hat{x}\hat{y}}\rangle=1$, we obtain
\[
\sum_{\hat{x}\in X}\sum_{\hat{y}\in Y}\langle u, m_{\hat{x}\hat{y}}\rangle\left[H\left(\{a_{x}(\omega_{\hat{x}\hat{y}})\}_{x}\right)+H\left(\{b_{y}(\omega_{\hat{x}\hat{y}})\}_{y}\right)\right]\ge \Gamma_{A, B},
\]
or equivalently (see \eqref{def:entropic MN}) 
\begin{align}
\label{eq:proof1}
H(A\mid\mathcal{M})+H(B\mid\mathcal{M})\ge \Gamma_{A, B}.
\end{align}
Because of the nonnegativity of (classical) conditional mutual information \cite{Cover:2006:EIT:1146355}, it holds that
\[
H(A\mid\mathcal{M}^{A})\ge H(A\mid\mathcal{M})\quad\mbox{and}\quad H(B\mid\mathcal{M}^{B})\ge H(B\mid\mathcal{M}),
\]
which proves the theorem together with \eqref{eq:proof1}.\qed
\end{pf}
Theorem \ref{thm:entropic MUR} is a generalization of the quantum result \cite{PhysRevLett.112.050401} to a class of GPTs. In fact, when we consider a finite dimensional quantum theory and a pair of rank-1 PVMs $A=\{\ketbra{a_{x}}{a_{x}}\}_{x}$ and $B=\{\ketbra{b_{y}}{b_{y}}\}_{y}$, our theorem results in the one in \cite{PhysRevLett.112.050401} with the quantum bound $\Gamma_{A, B}=-2\log\max_{x, y}|\braket{a_{x}|b_{y}}|$ by Maassen and Uffink \cite{PhysRevLett.60.1103}. Theorem \ref{thm:entropic MUR} demonstrates that if there is an entropic PUR, i.e. $\Gamma_{A, B}>0$, then there is also an entropic MUR which shows that we cannot make both $\mathsf{N}(\mathcal{M}^{A};A)$ and $\mathsf{N}(\mathcal{M}^{B};B)$ vanish.

\subsection{Examples: entropic uncertainty in regular polygon theories}
\label{subsec:eg UR}
In this part, we restrict ourselves to the regular polygon theories. Although the self-duality with respect to $\langle\cdot,\cdot\rangle_{G}$ holds only for regular polygons with odd sides, we can introduce the entropic measurement error \eqref{def:entropic MN} in the same way and prove the same theorem also for even-sided regular polygon theories. This is because, as shown in \cite{takakura2020uncertainty}, effects can be regarded as elements of $V_{+}$ and \eqref{eq:eigenstate} holds still in even-sided regular polygon theories by means of a suitable parameterization. We restate this fact as another theorem.
\begin{thm}
\label{thm:entropic MUR polygon}
Theorem \ref{thm:entropic MUR} also holds for the regular polygon theories.
That is, for a regular polygon theory with its state space $\Omega_{n}$ ($n=3, 4, \cdots, \infty$), ideal measurements $A=\{a_{x}\}_{x}$ and $B=\{b_{y}\}_{y}$ on $\Omega_{n}$, and an arbitrary approximate joint measurement $\mathcal{M}$ of $(A, B)$ with its marginals $\mathcal{M}^{A}$ and $\mathcal{M}^{B}$, 
if there exists a relation
\begin{align*}
H\left(\{a_{x}(\omega)\}_{x}\right)+H\left(\{b_{y}(\omega)\}_{y}\right)\ge \Gamma_{A, B}(n)\quad\ ^\forall\omega\in\Omega_{n},
\end{align*}
then
\begin{align*}
\mathsf{N}(\mathcal{M}^{A};A)+\mathsf{N}(\mathcal{M}^{B};B)\ge \Gamma_{A, B}(n).
\end{align*}
\end{thm}
In the following, we give a concrete value of $\Gamma_{A, B}(n)$ in Theorem \ref{thm:entropic MUR polygon} in the way introduced in subsection \ref{subsec:entropic PURs}.

Let us focus on the state space $\Omega_{n}$. Any nontrivial ideal measurement is of the form $\{e_{n}^{\mathrm{ext}}(i), u-e_{n}^{\mathrm{ext}}(i)\}$ (see \eqref{def:polygon pure effect} and \eqref{def:disc pure state and effect}). Thus, if we consider a pair of ideal measurements $A$ and $B$, then we can suppose that they are binary: $A=A^{i}\equiv\{a^{i}_{0}, a^{i}_{1}\}$ and $B=B^{j}\equiv\{b^{j}_{0}, b^{j}_{1}\}$ with $a^{i}_{0}=e_{n}^{\mathrm{ext}}(i)$ and $b^{j}_{0}=e_{n}^{\mathrm{ext}}(j)$ for $i, j\in\{0, 1, \cdots, n-1\}$ (or $i, j\in[0, 2\pi)$ when $n=\infty$). On the other hand, it holds that
\begin{equation}
	\begin{aligned}
		\label{eq:bound for states}
		\max_{x=0, 1}a^{i}_{x}(\omega)+\max_{y=0, 1}b^{j}_{y}(\omega)
		&\le\sup_{\omega\in\Omega_{n}}\max_{(x, y)\in\{0, 1\}^{2}}[(a^{i}_{x}+b^{j}_{y})(\omega)]\\
		&=\max_{\omega\in\Omega_{n}^{\mathrm{ext}}}\max_{(x, y)\in\{0, 1\}^{2}}[(a^{i}_{x}+b^{j}_{y})(\omega)]
	\end{aligned}
\end{equation}
because $\Omega_{n}$ is a compact set and any state can be represented as a convex combination of pure states. Therefore, if we let $\omega_{n}^{\mathrm{ext}}(k)$ be a pure state (\eqref{def:polygon pure state} and \eqref{def:disc pure state and effect}), then the value
\begin{align}
\label{eq:bound for pure state1}
\gamma_{A^{i}, B^{j}}:=\max_{k}\max_{(x, y)\in\{0, 1\}^{2}}[(a^{i}_{x}+b^{j}_{y})(\omega_{n}^{\mathrm{ext}}(k))]
\end{align}
gives a Landau-Pollak-type relation
\begin{equation}
\label{eq:LP bound}
\max_{x=0, 1}a^{i}_{x}(\omega)+\max_{y=0, 1}b^{j}_{y}(\omega)\le\gamma_{A^{i}, B^{j}}\quad\ ^\forall\omega\in\Omega_{n},
\end{equation}
to derive entropic relations
\begin{align}
H\left(\{a_{x}(\omega)\}_{x}\right)+H\left(\{b_{y}(\omega)\}_{y}\right)\ge -2\log\frac{\gamma_{A^{i}, B^{j}}}{2}\quad\ ^\forall\omega\in\Omega_{n},
\end{align}
and
\begin{align}
\mathsf{N}(\mathcal{M}^{A};A)+\mathsf{N}(\mathcal{M}^{B};B)\ge-2\log\frac{\gamma_{A^{i}, B^{j}}}{2}.
\end{align}
\begin{table}[p]
	\centering
	\caption{The value $(a^{i}_{x}+b^{j}_{y})(\omega_{n}^{\mathrm{ext}}(k))$ when $n$ is even.}
	\begin{tabular}{|c||c|}
		\hline
		$x=0, y=0$ & $1+r_{n}^{2}\cos\left[\frac{\theta_{i}+\theta_{j}}{2}-\phi_{k}\right]\cos\left[\frac{\theta_{i}-\theta_{j}}{2}\right]$              \rule[-3.5mm]{0mm}{10.5mm}                    \\ \hline
		$x=1, y=0$ & $1+r_{n}^{2}\sin\left[\frac{\theta_{i}+\theta_{j}}{2}-\phi_{k}\right]\sin\left[\frac{\theta_{i}-\theta_{j}}{2}\right]$                \rule[-3.5mm]{0mm}{10.5mm}                 \\ \hline
		$x=0, y=1$ & ($i \longleftrightarrow j$ in the case of $x=1, y=0$)                                                                          \rule[-3.5mm]{0mm}{10.5mm}    \\ \hline
		$x=1, y=1$ & $1-r_{n}^{2}\cos\left[\frac{\theta_{i}+\theta_{j}}{2}-\phi_{k}\right]\cos\left[\frac{\theta_{i}-\theta_{j}}{2}\right]$\rule[-3.5mm]{0mm}{10.5mm} \\ \hline
		\multicolumn{2}{|c|}{$\theta_{i}\equiv\frac{2i-1}{n}\pi,\quad\theta_{j}\equiv\frac{2j-1}{n}\pi,\quad\phi_{k}\equiv\frac{2k}{n}\pi$}  \rule[-3.5mm]{-1.3mm}{10.5mm} \\ \hline
	\end{tabular}
	\label{table:even}
\end{table}
\begin{table}[p]
	\centering
	\caption{The value $(a^{i}_{x}+b^{j}_{y})(\omega_{n}^{\mathrm{ext}}(k))$ when $n$ is odd.}
	\begin{tabular}{|c||c|}
		\hline
		$x=0, y=0$ & $\frac{2}{1+r_{n}^{2}}+\frac{2r_{n}^{2}}{1+r_{n}^{2}}\cos\left[\frac{\theta_{i}+\theta_{j}}{2}-\phi_{k}\right]\cos\left[\frac{\theta_{i}-\theta_{j}}{2}\right]$ \rule[-3.5mm]{0mm}{10.5mm}   \\ \hline
		$x=1, y=0$ & $1+\frac{2r_{n}^{2}}{1+r_{n}^{2}}\sin\left[\frac{\theta_{i}+\theta_{j}}{2}-\phi_{k}\right]\sin\left[\frac{\theta_{i}-\theta_{j}}{2}\right]$                    \rule[-3.5mm]{0mm}{10.5mm}    \\ \hline
		$x=0, y=1$ & ($i \longleftrightarrow j$ in the case of $x=1, y=0$)                                                                                                         \rule[-3.5mm]{0mm}{10.5mm}   \\ \hline
		$x=1, y=1$ & $\frac{2r_{n}^{2}}{1+r_{n}^{2}}-\frac{2r_{n}^{2}}{1+r_{n}^{2}}\cos\left[\frac{\theta_{i}+\theta_{j}}{2}-\phi_{k}\right]\cos\left[\frac{\theta_{i}-\theta_{j}}{2}\right]$    \rule[-3.5mm]{0mm}{10.5mm}    \\ \hline
		\multicolumn{2}{|c|}{$\theta_{i}\equiv\frac{2i}{n}\pi,\quad\theta_{j}\equiv\frac{2j}{n}\pi,\quad\phi_{k}\equiv\frac{2k}{n}\pi$}  \rule[-3.5mm]{-1.3mm}{10.5mm} \\ \hline
	\end{tabular}
	\label{table:odd}
\end{table}
\begin{table}[p]
	\centering
	\caption{The value $(a^{i}_{x}+b^{j}_{y})(\omega_{n}^{\mathrm{ext}}(k))$ when $n$ is $\infty$.}
	\begin{tabular}{|c||c|}
		\hline
		$x=0, y=0$ & $1+\cos\left[\frac{\theta_{i}+\theta_{j}}{2}-\phi_{k}\right]\cos\left[\frac{\theta_{i}-\theta_{j}}{2}\right]$                   \rule[-3.5mm]{0mm}{10.5mm}                       \\ \hline
		$x=1, y=0$ & $1+\sin\left[\frac{\theta_{i}+\theta_{j}}{2}-\phi_{k}\right]\sin\left[\frac{\theta_{i}-\theta_{j}}{2}\right]$                               \rule[-3.5mm]{0mm}{10.5mm}          \\ \hline
		$x=0, y=1$ & ($i \longleftrightarrow j$ in the case of $x=1, y=0$)                                                                                            \rule[-3.5mm]{0mm}{10.5mm}       \\ \hline
		$x=1, y=1$ & $1-\cos\left[\frac{\theta_{i}+\theta_{j}}{2}-\phi_{k}\right]\cos\left[\frac{\theta_{i}-\theta_{j}}{2}\right]$ \rule[-3.5mm]{0mm}{10.5mm}\\ \hline
		\multicolumn{2}{|c|}{$\theta_{i}\equiv i,\quad\theta_{j}\equiv j,\quad\phi_{k}\equiv k$}  \rule[-3.5mm]{-1.3mm}{10.5mm} \\ \hline
	\end{tabular}
	\label{table:infty}
\end{table}
Table \ref{table:even} - Table \ref{table:infty} show the value of $(a^{i}_{x}+b^{j}_{y})(\omega_{n}^{\mathrm{ext}}(k))$ in terms of the angles $\theta_{i}$, $\theta_{j}$, and $\phi_{k}$ between the $x$-axis and the effects $a^{i}_{0}=e_{i}^{\mathrm{ext}}(i)$, $b^{j}_{0}=e_{j}^{\mathrm{ext}}(j)$, and the state $\omega_{n}^{\mathrm{ext}}(k)$ respectively when viewed from the $z$-axis (see \eqref{def:polygon pure state0} - \eqref{def:disc pure state and effect}). As an illustration, let us consider the case when $n$ is a multiple of $4$ and $\theta_{i}-\theta_{j}=\frac{\pi}{2}$. From Table \ref{table:even}, we can calculate $\gamma_{A^{i}, B^{j}}$ and $\phi_{k}$ which gives the maximum in \eqref{eq:bound for pure state1}:
\begin{align}
\label{eq:LP in polygon 1}
\gamma_{A^{i}, B^{j}}=1+\frac{r_{n}^{2}}{\sqrt{2}}\qquad\left(\phi_{k}=\theta_{i}-\frac{\pi}{4}\right)
\end{align}
when $n\equiv 4\ (\mbox{mod}\ 8)$, and
\begin{align}
\label{eq:LP in polygon 2}
\gamma_{A^{i}, B^{j}}=1+\frac{1}{\sqrt{2}}\qquad\left(\phi_{k}=\theta_{i}-\frac{\pi}{4}\pm\frac{1}{n}\right)
\end{align}
when $n\equiv 0\ (\mbox{mod}\ 8)$. \eqref{eq:LP in polygon 1} and \eqref{eq:LP in polygon 2} are consistent with the case when $n=\infty$:
\begin{align}
\label{eq:LP in polygon 3}
\gamma_{A^{i}, B^{j}}=1+\frac{1}{\sqrt{2}}\qquad\left(\phi_{k}=\theta_{i}-\frac{\pi}{4}\right).
\end{align}
\eqref{eq:LP in polygon 3} for $n=\infty$ can be regarded as corresponding to the quantum result \eqref{eq:quantum LP} with $A$ and $B$ being, for example, the $X$ and $Z$ measurements on a single qubit respectively. Note that $\gamma_{A^{i}, B^{j}}$ can be used also to evaluate the nonlocality of the theory via its degree of incompatibility \cite{takakura2020uncertainty}.

\section{Conclusion and Discussion}
\label{sec:conclusion}
Overall, we examined entropic PURs and MURs in GPTs with transitivity and self-duality with respect to a specific inner product and in the regular polygon theories.  
We proved similar entropic relations to PURs and MURs in quantum theory also in the GPTs with the Landau-Pollak-type relations and the entropic measurement error generalized respectively.
It manifests that the entropic behaviors of two kinds of uncertainty in quantum theory are also observed in a broader class of physical theories, and thus they are more universal phenomena.
It is easy to obtain similar results if more than two measurements are considered.
We also gave concrete calculations of our results in the regular polygon theories.

The resulting theorems (Theorem \ref{thm:entropic MUR} and Theorem \ref{thm:entropic MUR polygon}) can be considered as entropic expressions of the ones in \cite{takakura2020uncertainty}.
Our theorems demonstrate in an entropic way that MURs are indicated by PURs and both of them can be evaluated by the same bound. We note similarly to  \cite{takakura2020uncertainty} that while the quantum results in \cite{PhysRevLett.112.050401} were based on the ``ricochet'' property of maximally entangled states, our theorems were obtained without considering entanglement or even composite systems. It may be indicated that some characteristics of quantum theory can be obtained without entanglement.

Although there are researches suggesting that our assumptions on theories are satisfied in the presence of several ``physical'' requirements \cite{1367-2630-19-4-043025,Barnum_2014,PhysRevLett.108.130401}, future study will need to investigate whether our theorems still hold in GPTs with weakened assumptions.
To give better bounds to our inequalities, and to find information-theoretic applications of our results are also future problems.


\section*{Acknowledgment}
TM acknowledges financial support from JSPS (KAKENHI Grant Number 20K03732).



\bibliographystyle{hieeetr} 
\bibliography{ref_entropicUR}

\end{document}